# Sensing with THz metamaterial absorbers

Longqing Cong and Ranjan Singh[*]

*Centre for Disruptive Photonic Technologies, Division of Physics and Applied Physics, School of Physical and Mathematical Sciences School of Physical and Mathematical Sciences, Nanyang Technological University, Singapore 637371, Singapore*

[*]Email: ranjans@ntu.edu.sg

**Abstract**

Metamaterial perfect absorbers from microwaves to optical part of the electromagnetic spectrum has been intensely studied for its ability to absorb electromagnetic radiation. Perfect absorption of light by metamaterials have opened up new opportunities for application oriented functionalities such as efficient sensors and emitters. We present an absorber based sensing scheme at the terahertz frequencies and discuss optimized designs to achieve high frequency and amplitude sensitivities. The major advantage of a perfect metamaterial absorber as a sensor is the sensitive shift in the absorber resonance frequency along with the sharp change in the amplitude of the resonance due to strong interaction of the analyte with the electric and the magnetic fields at resonant perfect absorption frequency. We compare the sensing performance of the perfect metamaterial absorber with its complementary structural design and planar metasurface with identical structure. The best FoM values obtained for the absorber sensor here is 2.67 which we found to be significantly higher than the identical planar metamaterial resonator design. We further show that the sensitivity of the sensor depends on the analyte thickness with the best sensitivity values obtained for thicknesses approaching $\lambda/4n$, with $\lambda$ being the free space resonance wavelength and $n$ being the refractive index of the analyte. Application of metamaterial absorbers as sensors in the terahertz spectral domain would be of tremendous significance due to several materials having unique spectral signature at the terahertz frequencies.

# 1. Introduction

The quest to bridge the terahertz (THz) gap in the electromagnetic spectrum has ushered enormous amount of research activities in the recent years [1-5]. With the nonionizing property of the so called "T-rays", the THz devices have attracted tremendous attention due to its broad applications in imaging, remote sensing, astronomical radiation detection, high resolution spectroscopy, and biomedical analytics [6-11]. In recent times, artificially designed metamaterial devices have emerged as an important tool to manipulate electromagnetic waves at the subwavelength scales due to its ease to engineer optical properties such as refractive index, permittivity, and permeability based on a periodic array of unit cells that is typically called "meta-atoms". Periodic collection of "meta-atoms" form the basis of many fascinating and exotic effects like negative refraction [12, 13], perfect lenses [14] and cloaking [15]. Metamaterials and plasmonic based devices have also shown to have excellent applications in chemical and biomedical sensing [16], surface enhanced spectroscopy [17], and near-field scanning optical microscopy [18]. The collective excitation of subwavelength metamaterial and plasmonic array structures lead to localized electric and magnetic resonances while interacting with the incident light which provides an excellent platform for electromagnetic sensing and manipulation [19-24].

Recently, there has been several works demonstrating sensing with planar metamaterial split ring resonators [25-39]. However, either the sensitivity or the figure of merit (FoM) of such planar metamaterial sensor is relatively lower due to lack of strong electromagnetic fields and the broad resonance linewidth. The lossy planar metamaterials does not allow higher FoM values, thereby arising the need to look beyond planar structures. Perfect metamaterial absorbers (PMAs) have emerged as strong candidates for absorbing electromagnetic waves [2, 40-52]. PMAs are typically three layered structure that consists of a micro/nanofabricated planar metasurface layer, a dielectric spacer followed by the ground plane in the propagation direction. It has been recently shown that these tri-layers in an absorber forms a Fabry-Perot like cavity and the absorption effect occurs due to the interference between the multiple reflection inside the cavity formed by the metasurface layer and the ground plane [46, 53, 54]. A similar asymmetric Fabry-Perot cavity based absorber was realized by Kats et al. in an extremely simple and ultrathin device that consisted of unpatterned dielectric layer on top of an opaque substrate [55]. A key feature of the Fabry-Perot cavity effect in the resonant absorbers implies that the light could be stored in such a cavity for several cycles of reflection between the two layers. Additionally, at perfect absorption resonance frequency, both strong electric and magnetic fields are excited. These two features of a PMA could be exploited for enhanced light matter interaction and would be ideal for sensing application as it was elegantly demonstrated at infrared frequencies first by Liu et al [41] and then by Cheng et al [43].

**2. Polarization insensitive perfectly symmetric absorber and its complementary design: Sensitivity analysis**

In this manuscript, we present a numerically simulated design and scheme of using a PMA for sensing applications at the technologically relevant THz frequencies. We chose a polarization insensitive design with 4-fold rotational symmetry and investigated the sensing capabilities of the THz absorber in terms of the change in resonance frequency and the amplitude depth of the reflection spectra. The maximum sensitivity achieved in this work is 163 GHz/RIU (Refractive Index Unit) and FoM value achieved is 2.67. Absorber based sensing at THz frequencies would have tremendous applications due to the unique spectral signature behavior of several explosives and organic materials in this domain.

We select a polarization insensitive cross shaped absorber (CSA) design used previously by several groups [43, 56, 57] and its complementary cross shaped absorber (CCSA) to investigate the sensing abilities of THz PMAs. As shown in Figs. 1 (a) and (c), we designed the parameters of CSA and CCSA unit cells with the length of cross shaped structure with $l$=130 μm, the thickness of aluminum ($\sigma$=3.56×e$^7$ S/m) as 200 nm for both the structured layer and the ground plane. The thickness of polyimide ($\varepsilon$=3.1+0.217$i$) spacer in CSA is taken as $h$=18 μm and for CCSA structure as $h$ = 31 μm in order to match the impedance for the perfect absorption. A silicon ($\varepsilon$=11.9+0.0476$i$) substrate with thickness of 500 μm is used. Periodic boundary condition is applied in numerical model.

With the incident electric field and wave vector direction as shown in Fig. 1, the excited electric fringing fields above the absorber surfaces at resonance are schematically illustrated. When the lossless analyte of different refractive indices is deposited on the top surface of CSA and CCSA, the change in electric and magnetic field distribution would lead to different frequency and amplitude response. We numerically calculate the response of CSA and CCSA using Finite-element Frequency-domain (FEFD) solver. The amplitude reflection spectra with different refractive indices are plotted in Figs. 1(b) and (d) with analyte thicknesses of 45 μm and 50 μm, respectively for the CSA and the CCSA structures. For the CSA sensing device shown in Fig. 1(b), the resonance frequency red shifts from 0.637 THz to 0.515 THz (shift of 122 GHz) and the reflection amplitude of the resonance changes sharply from 4.3% to 46.3% when the refractive index of analyte is varied from $n=1.0$ to $n=1.8$. For the CCSA structure in Fig. 1(d), the total frequency shift is about 130 GHz and the resonance amplitude change is 56.6% for the identical range of variation in the refractive index. The quality factor ($Q$) of the resonance reflection spectra for CSA is 7.036 and for CCSA is 7.189.

Based on the frequency shift and the amplitude modulation with the change in analyte refractive indices, we estimated the frequency sensitivity and the amplitude sensitivity of CSA and CCSA sensors. We looked at the reflection spectra and extracted the absorber resonance frequency and corresponding amplitude for a fixed thickness but varying refractive index of the analyte layer. The refractive indices varied from $n=1.0$ to $n=1.8$ in incremental steps of 0.1. The index dependent shift in resonance frequency and change in amplitude of the CSA sensor for two different analyte thicknesses of $t=5$ μm and $t=45$ μm are shown in Figs. 2(a) and 2(b). Linear fitting functions were used to fit the resonance shift and the amplitude change data. The frequency sensitivity is defined as $df/dn$, i.e. the slope of the linear fitting function, where $df$ represents the change in the resonance frequency and $dn$ represents the change in the refractive index. Amplitude sensitivity has been similarly defined as $dA/dn$ representing the change in the amplitude depth of the resonant absorber with the change in refractive index of the analyte. We calculate the sensitivity with different analyte thicknesses as shown in Figs. 2(a) and 2(b). It is observed that there is a very significant enhancement of frequency and amplitude sensitivities when the analyte thickness is gradually increased from $t=5$ μm to $t=45$ μm. The frequency sensitivity increases from 80 GHz/RIU to 152 GHz/RIU and the amplitude sensitivity increases from 7.4% $RIU^{-1}$ to 53.2% $RIU^{-1}$. Similar sensitivity enhancement is observed for the CCSA sensor when the analyte thickness is varied from $t$=5 μm to $t$=50 μm, as shown in Figs. 2(c) and 2(d).

We carried out detailed investigation of the analyte thickness dependent sensitivity of CSA and CCSA sensors as revealed in Figs. 3(a) and 3(b), respectively. The frequency sensitivity of both absorbers follow an exponential growth curve with the increasing analyte layer thickness as shown in Fig. 3. The maximum analyte thicknesses, at which CSA and CCSA frequency sensitivity saturates are 45 μm and 50 μm, respectively. It is worth noting that these analyte thicknesses with maximum sensitivities in both absorbers are close to $\lambda/4n$ value, with $\lambda$ being the absorption resonance wavelength and $n$ being the refractive index of the analyte. The amplitude sensitivity follows a linearly increasing trend for both of the absorber sensor design. However, this does not imply that the amplitude sensitivity continues to increase indefinitely. In fact, at a specific analyte thickness, the amplitude modulation saturates when either the refractive index reaches a high value or when the analyte thickness reaches a specific value at a particular analyte refractive index. For the case of a CCSA sensor, a similar trend is observed.

## 3. Sensing with electric and magnetic fields

In order to understand the underlying sensing mechanism of both CSA and CCSA sensors, we exhibit the simulated electric field distribution in the x-z plane and magnetic field distribution in the y-z plane in both of these absorber structures as shown in Fig. 4. It clearly reveals the spatial extent of the electric and magnetic field confined within the absorber structures as well as the fringing fields that extend above the surfaces of the CSA and CCSA absorbers that play a key role in sensing any dielectric that falls in the vicinity of these fields. Figure 4(a) shows the top view cross section of electric field distribution at $y=0$ cut-plane for CSA and Fig. 4(b) shows the top view at $y=7.5$ μm cut-plane for CCSA, where the centers of CSA and CCSA are defined as $x=0$ and $y=0$ in the coordinate system. Similarly in Figs. 4(c) and 4 (d), we show the magnetic field in the *y-z* plane for both the designs. The simulated electric and magnetic field distributions shown are at the impedance matched perfect absorption resonance frequency of CSA and CCSA sensors. From Fig. 4, we also observe that the fields are more tightly confined around the corners of the complementary metasurface structure in CCSA than the left and right edges of transverse strip of CSA. The extent of the spatial fringing field is also observed to be larger in case of CCSA, thus justifying the relatively larger analyte thicknesses at which CCSA sensitivity gets saturated as compared to the CSA sensor. The tightly confined field distribution determines the sensitivity of the sensing device and the spatial extent of the fringing field determines the largest thickness of analyte of a particular refractive index that could be sensed. Thus, we understand the explicit physical mechanism by looking at the field distribution in CSA and CCSA sensors to explain the difference in their sensing performance. In addition to the electric and magnetic field resonant enhancements in absorbers, we also exploit the multi pass light cycles within the Fabry-Perot cavity to realize a strong interaction between the matter to be sensed and the light that is stored via multi-reflection between the two metallic layers separated by the dielectric spacer in the absorber cavity.

## 4. Figure of Merit (FoM) comparison between absorbers and planar metasurfaces

For a quantitative description of the sensing performance of the absorbers, we estimated the FoM which is typically defined as FoM = Sensitivity/FWHM, where FWHM is full width half maximum of the amplitude resonance of the sensing device. Here, we mainly focus on the frequency sensitivity of CSA and CCSA sensors. Based on the frequency sensitivity for several different thicknesses of analyte overlayer with different refractive indices as shown in Fig. 3, we calculated the FoM for both CSA and CCSA sensors, where we averaged the FWHM of the absorber resonance for gradually increasing refractive indices (ranging from $n = 1$ to $n = 1.8$). The FoM of CSA and CCSA with varying analyte thickness is plotted in Figs. 5(a) and 5(b) respectively. With CSA sensor the best FoM of 2.67 is observed for optimal analyte thickness of 20 μm and in CCSA, the best FoM of 2.05 is observed at analyte thickness of 25 μm. The FoM values obtained by using both types of absorbers presented here are significantly higher than planar metamaterial resonators as we would discuss subsequently.

In order to compare the performance of an absorber sensor and a simple single layer planar metasurface, we also simulated the sensitivity and FoM values of the corresponding planar metasurface top layer of CSA and CCSA by eliminating the ground metal plane from the two layer absorber structures. The FoM values of the corresponding planar metasurfaces are shown in Figs. 5(a) and 5(b) and it could be easily noticed that the values are significant lower than the absorber counterparts. The detail performance comparison data in terms of $Q$ factor and FoM are shown in Table 1.

## 5. Dielectric spacer thickness dependent sensitivity

With significantly better performance in terms of FoM, the MPAs have been shown to be excellent candidates for sensing application. However, since the thickness of the dielectric spacer layer is key in impedance matching of absorber with free space, we probed the sensing performance of the absorber with the change in the spacer thickness as this thickness may vary when a real sensing device is fabricated in a cleanroom environment. We calculated and analyzed the absorber sensitivity versus spacer thickness for both absorber designs, CSA and CCSA. We fixed the analyte thickness at $t$=20 μm for CSA sensor and $t$=25 μm for the CCSA sensor since the best FoMs were observed for these specific analyte thicknesses in each of the absorber sensors with corresponding spacer thicknesses of $h$=18 μm in CSA and $h$=31 μm in CCSA sensors. The frequency sensitivity of CSA sensor as shown in Fig. 6(a) does not change much beyond the spacer thickness of about 15 μm which demonstrates good stability of the CSA sensing device. However, the FoM in Fig. 6(a) shows a drastic decline as the spacer thickness increases beyond 15 μm. This occurs due to the broadening of the absorber resonance when the impedance is mismatched. Also, according to the interference effect in PMAs, the spacer thickness determines the constructive or destructive interference between the reflected light waves in the Fabry-Perot cavity of the absorber [46]. Constructive interference in a cavity occurs at specific wavelengths where the phase change of a round trip is $2\beta_r(\omega_0)+\Phi_{21}(\omega_0)+180°\approx360°$ with $\beta_r(\omega_0)$ as the propagation phase in spacer and $\Phi_{21}(\omega_0)$ as the reflection phase response of top metasurface. A sharp dispersive response of $\Phi_{21}(\omega)$ will lead to a dramatic broadening of absorber resonance line profile when the spacer thickness is changed from optimum, i.e. the change of $\beta_r(\omega)$, due to the deviation from the constructive interference condition. It is also important to note here that the thickness of THz metamaterial perfect absorber is much smaller than the conventional $\lambda/4$ Fabry-Perot cavity. We observe that in CCSA sensor, the sensitivity and FoM obtained saturation and remains stable with spacer thicknesses higher than 30 μm. Thus, we would like to stress that CCSA sensor design is more robust than the CSA design from the viewpoint of the reliability of the fabricated device with specific spacer thickness, although the FoM is slightly lower than that of the CSA sensor.

## 6. Conclusion

In conclusion, we present two different engineered designs of using a perfect metamaterial absorber as a sensor in the THz regime. We studied the detailed sensitivity aspect of the absorber designs that is required to get the best sensing performance. In comparison to the single layer planar metamaterials, the two designs of absorbers presented here shows significantly higher FoM values. Since a PMA behaves like a Fabry-Perot cavity, we exploit the enhanced electromagnetic fields and multi cycle light path within the cavity to interact strongly with the analyte. This aspect becomes the key reason for using absorber as a sensing device. An absorber as a sensor would be an important addition to the device starved THz regime that holds significance due its finger print spectral range for several explosive and organic materials.

**Acknowledgements**

The authors thank R. Yahiaoui, G. Dayal, and D. R. Chowdhury for fruitful discussions.

This work was funded by NTU startup grant.

**Fig. 1**

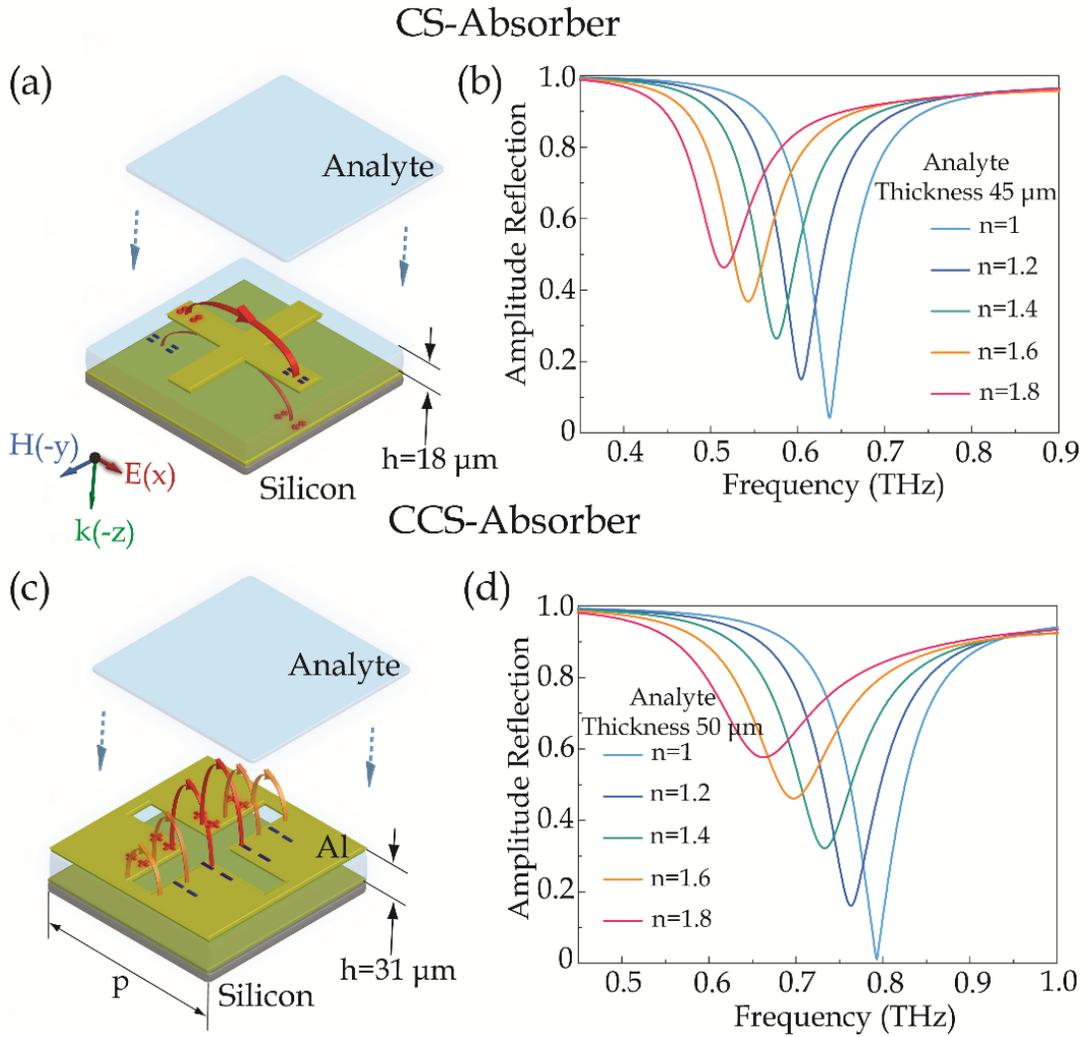

FIG. 1. (a) Cross shaped absorber (CSA) design for terahertz sensing (b) Amplitude reflection spectra of 45 μm thick analyte overlayer with varying refractive index (c) Complementary cross shape absorber (CCSA) design model for terahertz sensing (d) Amplitude reflection spectra of 50 μm thick analyte overlayer with varying refractive index. Periodicity of CSA and CCSA is 150 μm by 150 μm. The cross structure has arm width of 15 μm and length of 130 μm.

**Fig. 2**

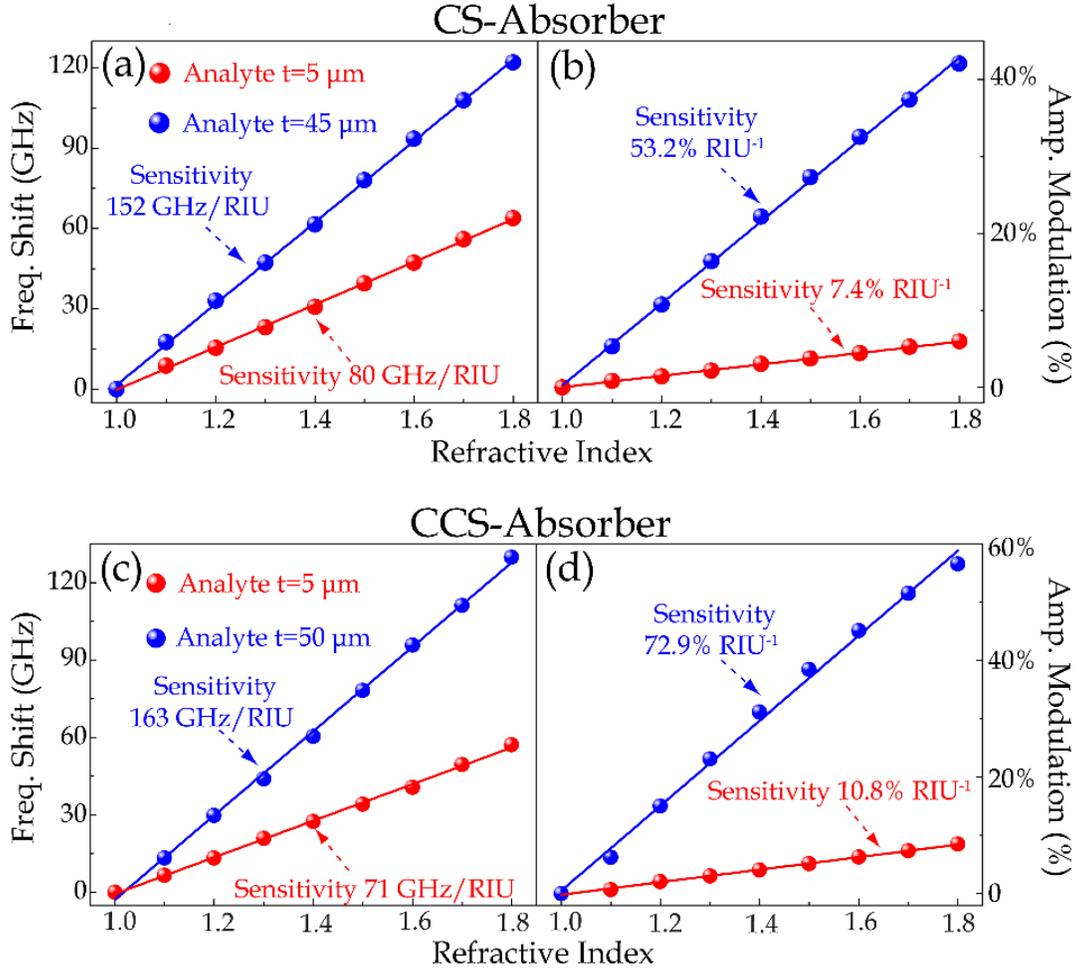

FIG. 2. (a) The resonance frequency shift and (b) reflection amplitude spectra modulation of cross shaped absorber (CSA) with varying analyte refractive index at analyte thicknesses of 5 µm and 45 µm, respectively. (c) The resonance frequency shift (d) and reflection amplitude spectra modulation of complementary cross shaped absorber (CCSA) with varying analyte refractive index at thicknesses of 5µm and 50 µm, respectively. (The solid lines are the linear fit for extracting the sensitivity).

**Fig. 3**

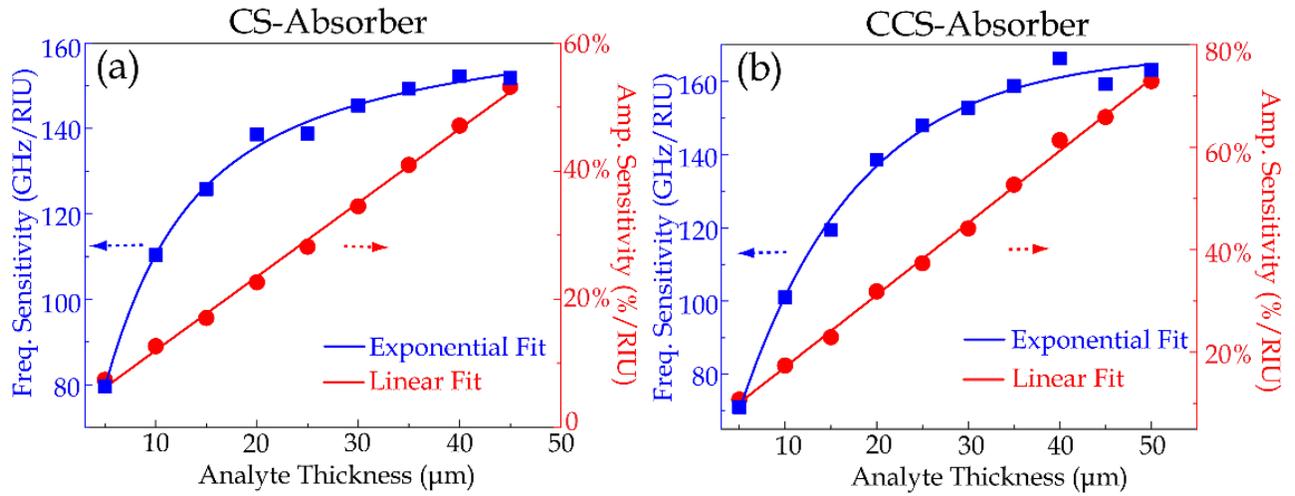

FIG. 3. Frequency and amplitude sensitivity versus analyte overlayer thickness for (a) CSA and (b) CCSA. (The solid blue lines are the exponential fit for the frequency sensitivity and the solid red lines are the linear fit for the amplitude sensitivity).

**Fig. 4**

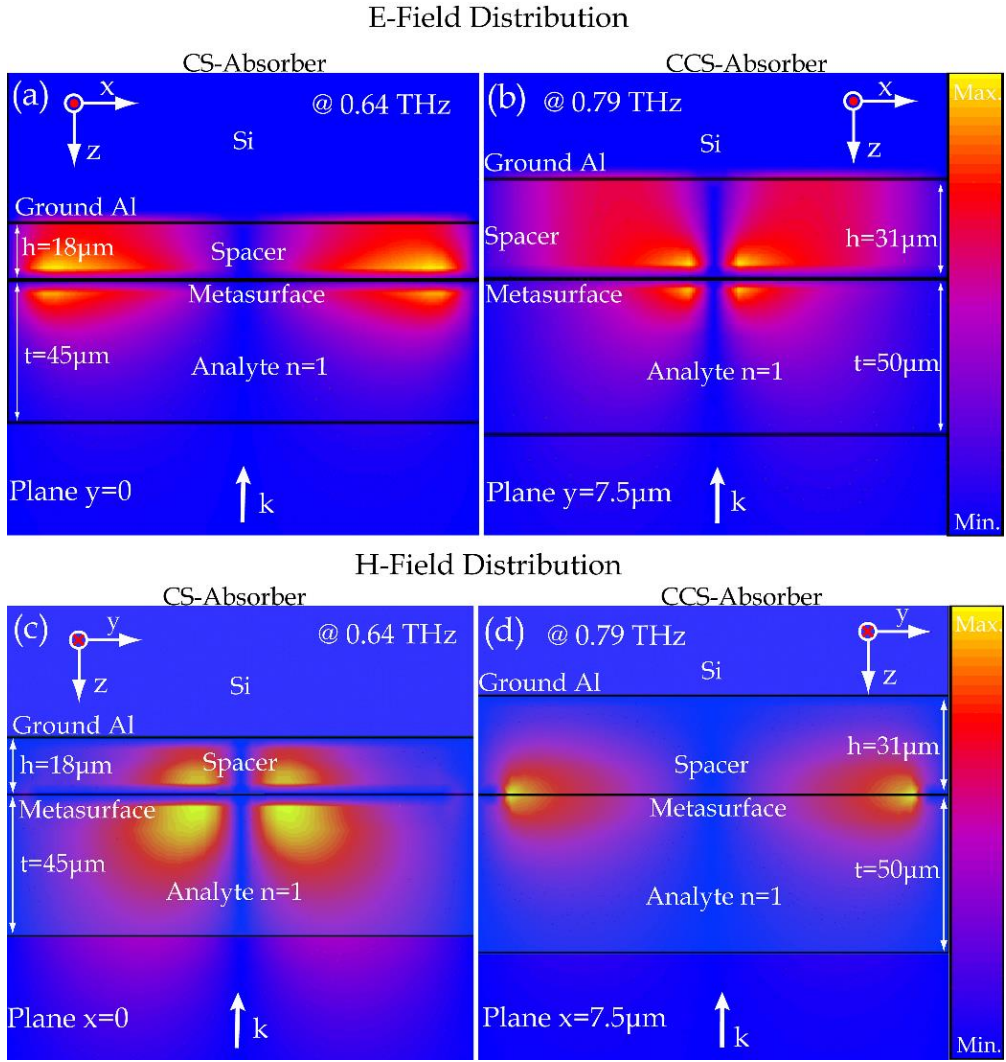

FIG. 4. Simulated electric field in z-direction (a) at cross section of *y*=0 cut-plane in CSA at resonance frequency of 0.64 THz and (b) at cross section of *y*=7.5 µm cut-plane in CCSA at resonance frequency 0.79 THz. Magnetic field distribution in *z*-direction (c) at cross section *x*=0 cut plane in CSA at 0.64 THz and (d) at cross section *x*=7.5 µm cut plane in CCSA at 0.79 THz.

**Fig. 5**

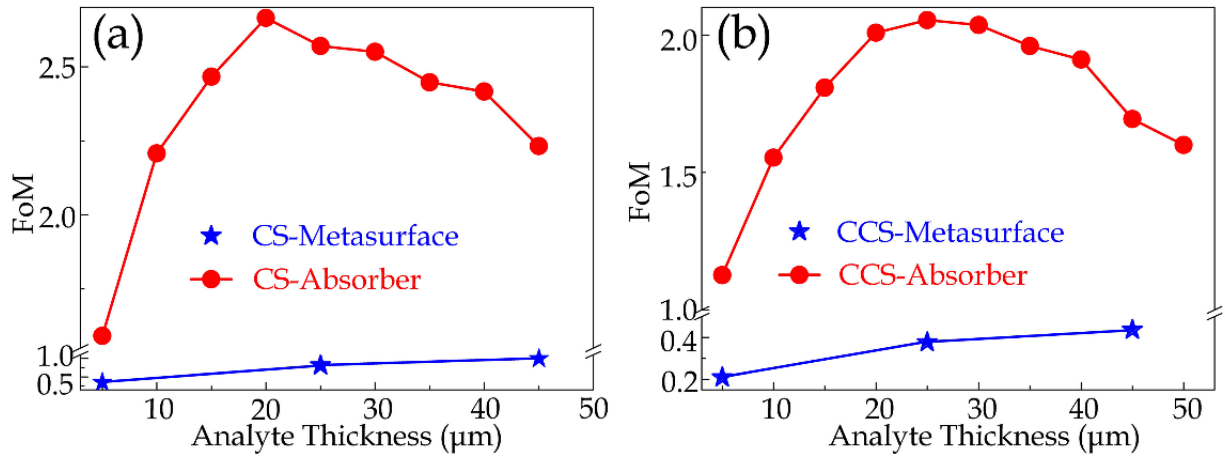

FIG. 5. (a) FoM of CSA (red circles) and the corresponding single layer planar metasurface (blue stars) versus analyte overlayer thickness. (b) FoM of CCSA (red circles) and the corresponding single layer planar metasurface (blue stars) versus analyte overlayer thickness.

**Fig. 6**

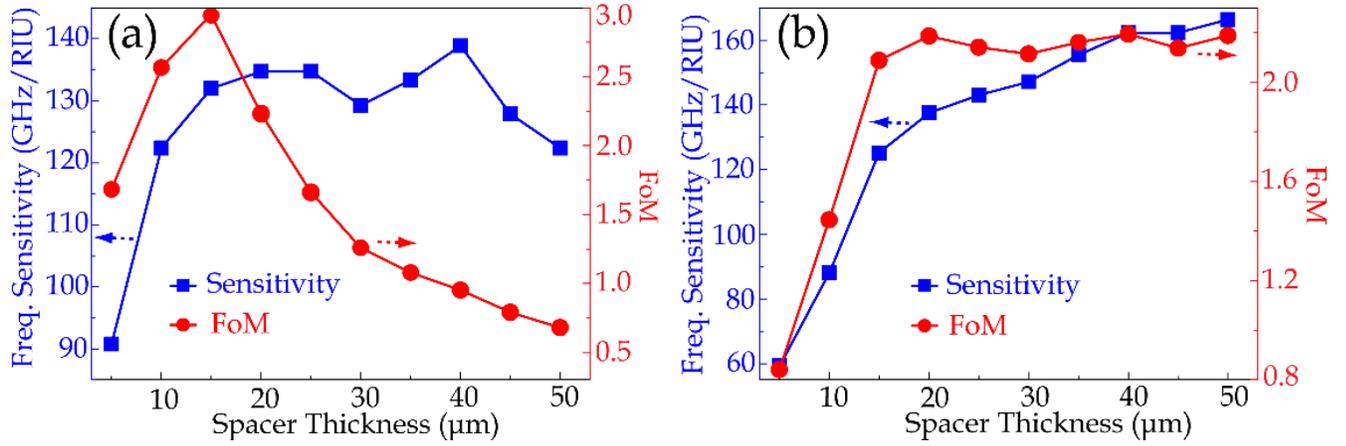

FIG. 6. (a) Frequency sensitivity and FoM of CSA versus spacer thickness with analyte overlayer thickness fixed at 20 μm. (b) Frequency sensitivity and FoM of CCSA versus spacer thickness with analyte overlayer thickness fixed at 25 μm.

**Table 1. Performance comparison**

|  | CS-Absorber | CCS-Absorber | CS-Metasurface | CCS-Metasurface |
|---|---|---|---|---|
| Q factor | 7.036 | 7.189 | 1.311 | 3.245 |
| FoM | 2.67 | 2.05 | 0.82 | 0.38 |